\documentclass[
reprint,
 amsmath,amssymb,
 aps, nofootinbib
]{revtex4-2}

\usepackage{graphicx}
\usepackage{dcolumn}
\usepackage{bm}

\begin{document}

\preprint{}

\title{Reconstruction wedges in AdS/CFT with boundary fractal-like structures}

\author{Ning Bao}
\email{ningbao@bnl.gov}
\affiliation{Computational Science Initiative, Brookhaven
National Laboratory, Upton, NY 11973 USA}

\author{Joydeep Naskar}
\email{naskar.j@northeastern.edu}
\affiliation{Department of Physics, Northeastern University, Boston, MA 02115, USA}

\date{\today}

\begin{abstract}
In this work, we show the robustness of uberholography and its associated quantum error correcting code against the breakdown of entanglement wedge in the presence of highly entropic mixed states in the bulk. We show that for Cantor-set-like erasure in the boundary in $AdS_3/CFT_2$, the code distance is independent of the mixed-state entropy in the bulk in the $m\rightarrow\infty$ limit. We also show that for a Sierpinski triangle shaped boundary subregion with fractal boundary erasures in $AdS_4/CFT_3$, bulk reconstruction is possible in the presence of highly entropic mixed states in the bulk in the large $m$ regime.
\end{abstract}

\maketitle


\section{\label{sec:level1}Introduction}

The study of quantum error correction has proven quite fruitful in the context of AdS/CFT, from the resolution of the commutator puzzle \cite{harlow2015} to the formalization of bulk reconstruction \cite{happycode2015}\cite{dong2016}\cite{hayden2016}\cite{faulkner2014} to ideas on subregion duality \cite{jlms2016}.  For recent reviews on holography and quantum error correction, see \cite{ayan2021-ref-request}\cite{eisert2021}.

Increasingly, there is also interest in how holographic studies of quantum error correction can influence quantum error correction as applied in quantum computing. There are some known limitations of existing non-holographic quantum error correcting codes such as the surface code, particularly in terms of their robustness against fractal noise \cite{arpit2021}. In a previous work\cite{bao2022}, we demonstrated the robustness of the holographic code and presumably codes derived thereupon to such types of noise using entanglement wedge-based arguments.

In the present work, we seek to expand upon this study by studying robustness properties against fractal noise in the presence of a black hole background in AdS/CFT, or equivalently at states of finite temperature in the boundary CFT. The work will be divided as follows. In sections 2 and 3 we will give a brief review of state-specific bulk-reconstruction and uberholography, as developed in \cite{akers-reconstruction-wedge-2019} and \cite{pastwaski2016}, respectively. In section 4, we will generalize this study to higher dimensions, in the context of a black hole background, and demonstrate that the robustness property continues to hold. Finally, we will give some concluding comments in section 5.

\section{State-specific reconstruction}
Recently it has been shown in \cite{akers-reconstruction-wedge-2019} that there is a macroscopic breakdown of the entanglement wedge when the von Neumann entropy of an object in the bulk cannot be ignored. In particular, this can occur when the bulk state is a highly mixed state. While one way to get past this is to purify those highly mixed states into e.g Bell pairs, the existence of this breakdown nevertheless establishes the fact that the entanglement wedge is state-specific in general. An object that was defined in the study of such breakdowns is the reconstruction wedge, defined by \cite{akers-reconstruction-wedge-2019} as the following:

\textbf{Definition:} The \textit{reconstruction wedge} of a boundary region A is the intersection of all entanglement wedges of
A for every state in the code subspace, pure or mixed. It is the region of space-time within which bulk operators are guaranteed to be reconstructible from the boundary in a state-independent manner.

Recently, it has also become clear that the entropy of the boundary at higher orders is no longer given by the RT surface area, but a more generalized entropy is associated with the quantum extremal surface, which is also defined in a state-dependent way\cite{wall-extremal-surface-2015}. More recent works on quantum error correction and quantum extremal surfaces have further established the importance of state-dependent bulk reconstruction\cite{akers-qms-2022}. Non-isometric codes\cite{harlow-akers} seem to suggest some state dependence in the non-isometric encoding in the sense of not modifying simple states but heavily modifying complex states.

The key idea is to relate the bulk entropy with the boundary entropy while taking into account the entropy of the bulk state. For a boundary subregion B corresponding to a bulk subregion \textbf{b} with a bulk state $\rho$ with von Neumann entropy \textbf{$S_b(\rho)$} the entanglement entropy $S(B)_{\rho}$ is given by \cite{faulkner2013}
\begin{equation}
    S(B)_{\rho}=\frac{A_B(\textbf{b})}{4G}+\textbf{$S_b(\rho)$}
\end{equation}
where G is Newton's constant in 2+1 dimensions and $A_B(\textbf{b})$ is the area of the bulk surface bounding \textbf{b}.

\section{\label{sec:level2}Uberholography in $AdS_3/CFT_2$}
The scheme of uberholography in $AdS_3/CFT_2$ was first proposed in \cite{pastwaski2016} and was extended to $AdS_4/CFT_3$ in \cite{bao2022}. More literature related to uberholography can be found in \cite{chen2020}\cite{kang2020}\cite{uberholo2022}.

\begin{figure}[h!]
		\centering
		\includegraphics[width=0.5\textwidth]{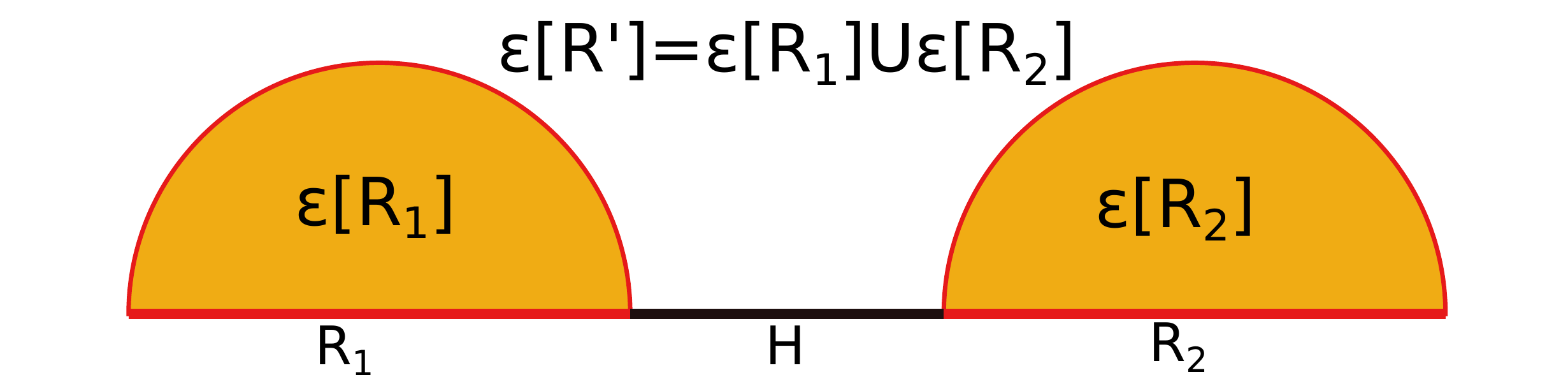}
	\caption{Surface $\chi_{R'}=\chi_{R1}\cup\chi_{R2}$. Disconnected regions $R_1$ and $R_2$ are drawn in red. The hole $H$ is drawn in black. The shaded region is the entanglement wedge $\epsilon[R']=\epsilon[R_1]\cup \epsilon[R_2]$.}
	\label{fig:1dfractcal-example1.1}
\end{figure}

\begin{figure}[h!]
		\centering
		\includegraphics[width=0.5\textwidth]{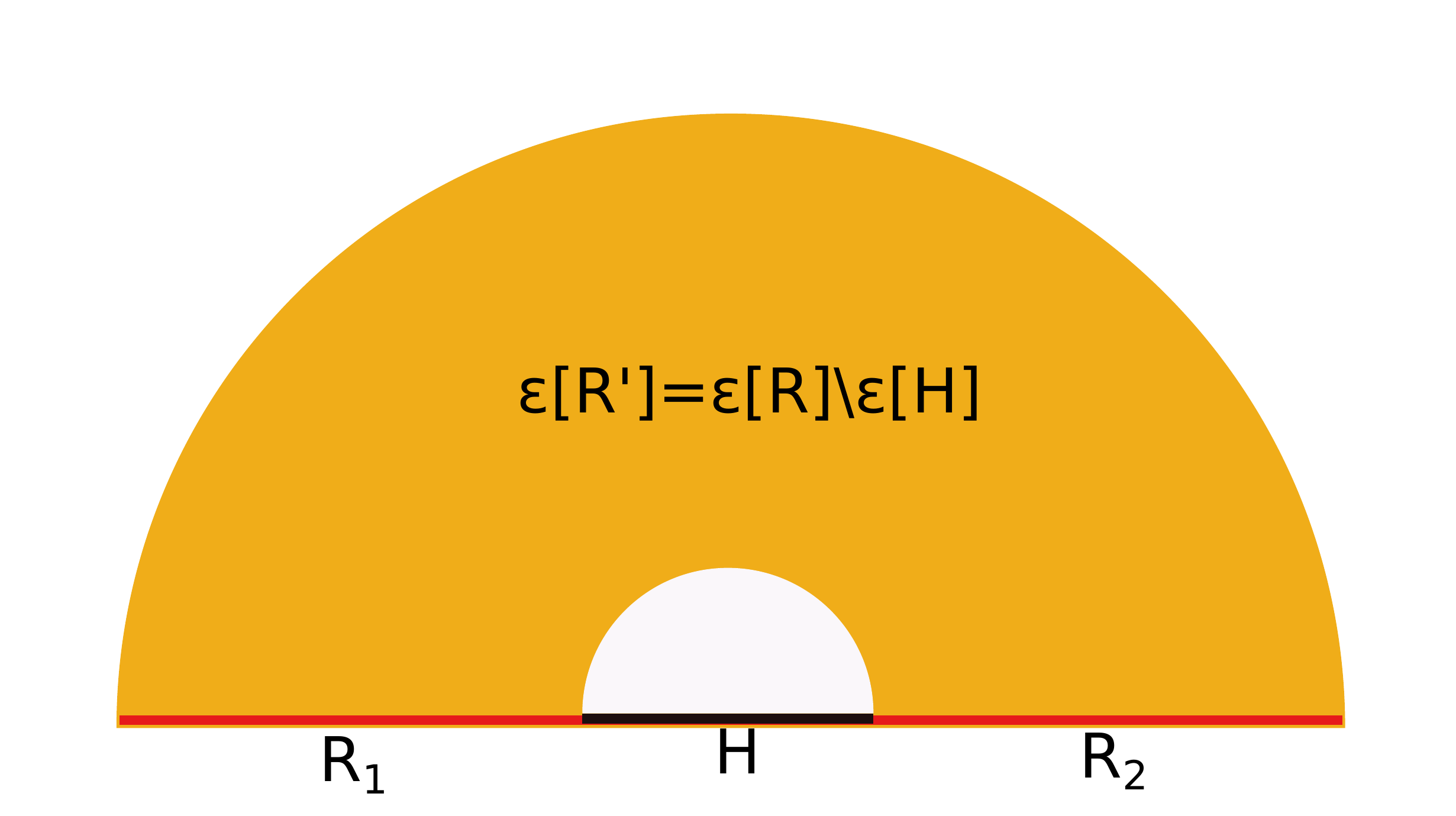}
	\caption{Surface $\chi_{R'}=\chi_{R}\cup\chi_{H}$. Disconnected regions $R_1$ and $R_2$ are drawn in red. The hole $H$ is drawn in black. The shaded region is the entanglement wedge $\epsilon[R']=\epsilon[R]\backslash \epsilon[H]$.}
	\label{fig:1dfractcal-example1.2}
\end{figure}

The idea of \emph{uberholography} is as follows is: Given a boundary subregion $R$, one punches a hole $H$ from it leaving disjoint boundary subregions $R_1$, $R_2$ and hole $H$ as shown in figures \ref{fig:1dfractcal-example1.1}, \ref{fig:1dfractcal-example1.2} with their respective entanglement wedges. The hole $H$ is punched such that with $0<r<1$
    \begin{equation}
    |R_1|=|R_2|=(\frac{r}{2})|R|, \quad |H|=(1-r)|R|.
\end{equation}
For the same boundary subregion $R$, there are two candidate surfaces that might be the minimal surface, giving rise to two different candidate entanglement wedges. Bulk reconstruction is possible when the RT surface (in this dimension, a geodesic) corresponding to the candidate entanglement wedge $\epsilon[R']=\epsilon[R] \backslash \epsilon[H]$(figure \ref{fig:1dfractcal-example1.2}) is smaller than the one corresponding to $\epsilon[R']=\epsilon[R_1] \cup \epsilon[R_2]$(figure \ref{fig:1dfractcal-example1.1}) where $\epsilon[A]$ is the entanglement wedge in the bulk corresponding to boundary subregion $A$ and the RT surface of this entanglement wedge is denoted by $\chi_A$. The area of this RT surface is simply given by \cite{cardy-calabrese}\cite{brown-henneaux}
\begin{equation}
    |\chi_A| = 2L \log{\left(\frac{|A|}{a}\right)},
\end{equation}
where $L$ is the AdS radius and $|A|$ denotes the length of $A$.
The division of the boundary subregion $R$ into $H$, $R_1$ and $R_2$ imposes a constraint on the allowed values of $r$ as it has to satisfy
\begin{equation}
    |\chi_{R_1}|+|\chi_{R_2}| \geq |\chi_{R}|+ |\chi_{H}|
\end{equation}
at level 1\footnote{Level 0 indicates no holes in the boundary, Level 1 refers to punching out holes once, Level 2 twice and so on...}. Solving the critical case saturating this inequality, one arrives at the value 
\begin{equation}
\label{2d_r_crit_value}
    r_C=2(\sqrt{2}-1),
\end{equation} where $r_C$ is the critical value of $r$ for the continuous phase to be favoured over the discontinuous one. We will drop the subscript C from now on for convenience and just write $r$.
It is interesting to note here that as we increase the number of iterations, the value of this critical $r$ remains unchanged. This property of $m$-independence breaks down by the inclusion of bulk entropy \textbf{$S_b$}\footnote{The bulk entropy \textbf{$S_b$} could be arising from a pure or mixed state. We shall be referring it as the entropy of the maximally mixed state.}.

Recalling the definition of distance of the code with operator algebra $A$ in bulk region $X$ to be 
\begin{equation}
d(A_X)\leq \frac{|R_{min}|}{a}=2^m=\left(\frac{|R|}{a}\right)^{\alpha},
\end{equation}
where
\begin{equation}
    \alpha=\frac{\log{2}}{\log{2/r}}=\frac{1}{\log_2{(\sqrt{2}+1})}=0.786
\end{equation}
and $R_{min}$ is the union of boundary subregions remaining after punching fractals holes until level $m$ when the size of the smallest subregion is of the order of short-distance cutoff '$a$'.

\subsection{Reconstruction Wedge in $AdS_3/CFT_2$}

There are two questions that we will address in this section. First, does uberholography survive state-dependent reconstruction, i.e, can we still perform bulk reconstruction with fractal erasures in the boundary, when the bulk has a mixed state with large von Neumann entropy? Second, even if the scheme of reconstructing bulk information with fractal erasures on the boundary survives, to reconstruct the bulk information, how does the support on the boundary change, i.e, does it increase with the introduction of a maximally mixed state in the bulk? We find that the answer to the first question is yes at every level of iteration and the answer to second question turns out to be "no" to leading order as $m\rightarrow\infty$. The effect of entropy of the bulk state is exponentially suppressed with increasing iteration.

\begin{figure}[h!]
		\centering
		\includegraphics[width=0.5\textwidth]{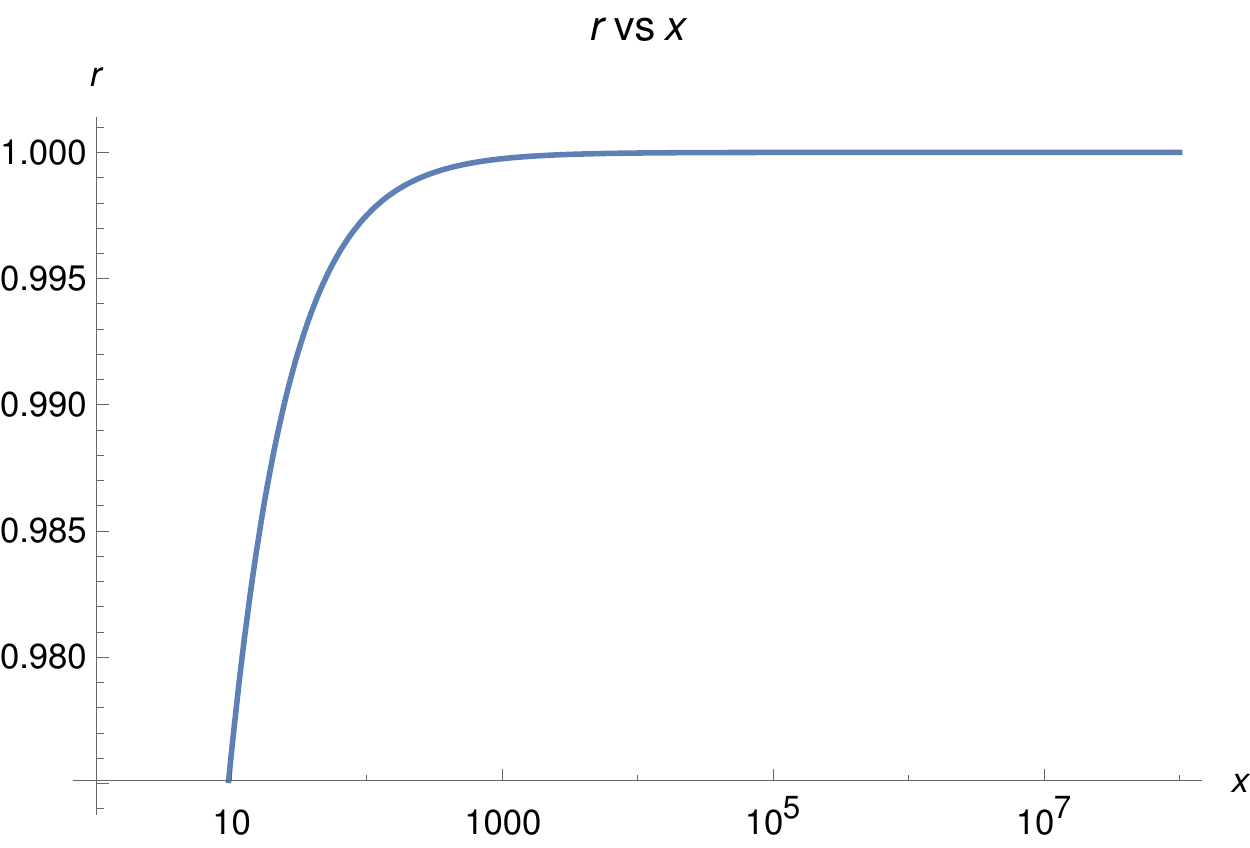}
	\caption{r vs x plot at level 1. One can always tune $0<r<1$, such that regardless of how large \textbf{$S_b$} is, uberholographic error correction is always possible.}
	\label{fig:level1-rvsx}
\end{figure}

Consider a state in the bulk with entropy \textbf{$S_b$}. At the first iteration, for the continuous phase to dominate over the discontinuous one, the following inequality must be satisfied:
\begin{equation}
    |\chi_{R_1}|+|\chi_{R_2}| \geq |\chi_{R}|+ |\chi_{H}| + 4G\textbf{$S_b$}
\end{equation}
hinting an upper bound on \textbf{$S_b$} for a scheme for bulk reconstruction of deep interior operators to exist. This perceived upper bound is given by
\begin{equation}
    4G\textbf{$S_b$}\leq 2L \log\left({\frac{(r/2)^2}{(1-r)}}\right).
    \label{leve1Sb}
\end{equation}
Re-arranging, the critical condition for $r$ is given by
\begin{equation}
\frac{\left(\frac{r}{2}\right)^2}{(1-r)}=e^{4G\textbf{$S_b$}/2L}.
\end{equation}
Setting $\textbf{$S_b$}=0$, we recover the condition in \cite{pastwaski2016}
\begin{equation}
   \frac{\left(\frac{r}{2}\right)^2}{(1-r)}=1,
\label{pastwaski-condition}    
\end{equation}
giving us (\ref{2d_r_crit_value}).
Apparently it may look like from (\ref{leve1Sb}) that at level 1, \textbf{$S_b$} is bounded above by a finite value and as the bulk entropy becomes larger than this value, the connected phase gets dominated by the disconnected one, implying that information in the bulk can no longer be reconstructed. But it is not the case, as we will show here. For simplicity, let $x=e^{4G\textbf{$S_b$}/2L}$, which gives us the inequality
\begin{equation}
    r^2+4rx-4x \geq 0.
\end{equation}
Solving for the critical case of equality, we get
\begin{equation}
    r=2x\left(\sqrt{1+\frac{1}{x}}-1\right).
\end{equation}
Setting $x=1$($\textbf{$S_b$}=0$), we recover the value of $r$ as in (\ref{2d_r_crit_value}). But it is clear that $0<r<1$ for any value of \textbf{$S_b$}, making bulk reconstruction possible in principle. However, as we increase \textbf{$S_b$}, $r$ approaches 1, making it less viable for any practical considerations (see fig \ref{fig:level1-rvsx}).

\begin{figure}[h!]
		\centering
		\includegraphics[width=0.5\textwidth]{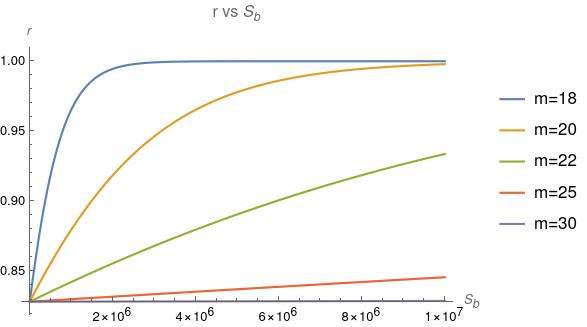}
	\caption{$r$ vs $S_b$ plot for various $m$. At $m\rightarrow\infty$, the curve is given by the line $r=2(\sqrt{2}-1)$. We have set $4G/L=1$ for simplicity.}
	\label{fig:vary-m}
\end{figure}

Our central interest is in understanding the behaviour at large $m$. At level $m$ of the fractal boundary erasure, the disconnected phase has a RT surface given by:
\begin{equation}
    |\chi_{R'}|_{disconn.}= 2L \left[2^m \log\left({\frac{(\frac{r}{2})^m |R|}{a}}\right) \right],
\end{equation}
while the connected phase has a RT surface
\begin{equation}
\begin{aligned}
    |\chi_{R'}|_{conn.}=\\
    2L\left[ \log\left(\frac{|R|}{a}\right)+
    \sum_{j=1}^m 2^{j-1} \log\left(\frac{ \left(\frac{r}{2}\right)^{j-1}(1-r)|R|}{a} \right) \right].
\end{aligned}
\end{equation}
With a bulk state having entropy \textbf{$S_b$}, we arrive at the following inequality in place of (\ref{leve1Sb}):
\begin{equation}
    4G\textbf{$S_b$}\leq 2L (2^m-1) \log\left({\frac{(r/2)^2}{(1-r)}}\right).
    \label{levelmSb}
\end{equation}
In the $m\rightarrow\infty$ limit, bulk reconstructability holds for any finite non-zero bulk entropy \textbf{$S_b$}, i.e, the connected phase always dominates over the disconnected one and the scheme of quantum error correction via uberholography for fractal erasures on the boundary is robust against state-dependent reconstruction.
However, the support on the boundary increases. We will now show that this increment is very small (negligible). Inverting equation (\ref{levelmSb}) and considering the critical equality, we get the condition on r

\begin{equation}
    \frac{\left(\frac{r}{2}\right)^2}{(1-r)}=e^{4G\textbf{$S_b$}/2L(2^m-1)}.
\end{equation}
In the limit $m\rightarrow\infty$, the $RHS\rightarrow 1$, approaching the value of $r$ in $\textbf{$S_b$}=0$ limit \cite{pastwaski2016}
\begin{equation}
    r\rightarrow 2(\sqrt{2}-1).
    \label{level-m-infinity-r}
\end{equation}
Comparing this result with figure \ref{fig:level1-rvsx}, the asymptotic value of $r$ at a higher level $m\rightarrow\infty$ is not 1, but (\ref{level-m-infinity-r}) (see Fig. \ref{fig:vary-m}).

Thus, the code distance remains the same and we have a robust, state-independent uberholographic quantum error correcting code.

\begin{figure}[h!]
		\centering
		\includegraphics[width=0.48\textwidth]{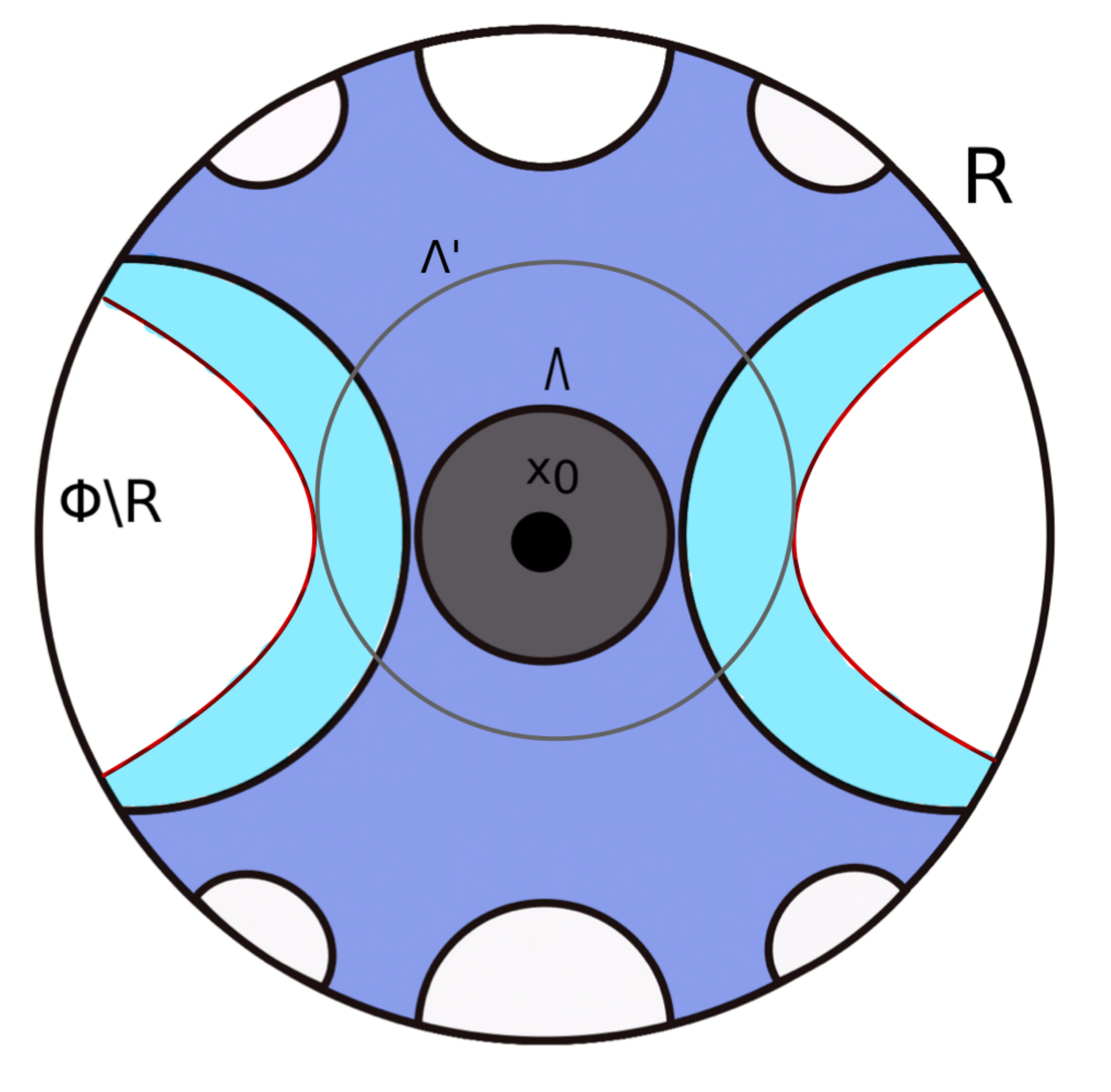}
	\caption{The inner logical boundary $\Lambda$ is contained inside the entanglement wedge (shaded in blue) of outer boundary subregion $R\subset \Phi$ where $\Phi$ is the outer physical boundary. We keep punching more holes of decreasing size in the boundary such that the remaining boundary region $R$(whose measure goes to zero when $m\rightarrow \infty)$ has an entanglement wedge that contains the logical boundary. By introducing a bulk mixed-state with large entropy, the volume of the entanglement wedge increases as the support on the outer boundary $\Phi$ increases. The new minimal area at level 1 is now the red line which does not touch the inner boundary $\Lambda$. One may imagine it as touching a larger inner boundary $\Lambda'$.}
	\label{fig:reconstruction_wedge}
\end{figure}

This becomes more clear in the context of the construction discussed in \cite{pastwaski2016}. In figure \ref{fig:reconstruction_wedge}, with $\textbf{$S_b$}=0$, $\Lambda$ is the logical boundary and $\Phi$ is the physical boundary. The boundary subregion R is the remaining boundary after punching out holes. One may interpret the introduction of a bulk state with entropy \textbf{$S_b$} as the bulk geodesic moving further away from the logical inner boundary $\Lambda$ increasing the support on the boundary for bulk reconstruction. One may imagine a boundary $\Lambda'$ to be the new "effective" logical boundary with the inclusion of a mixed state in the bulk. This means that the radius of the inner boundary $r_{in}$ gets modified to some $r'_{in}$ ($r'_{in}>r_{in}$) while the radius of the outer boundary $r_{out}$ remains the same. As long as $r_{out}$ is much greater than $r'_{in}$, the bulk reconstruction is possible. This is a reasonable assumption as $r_{out}$ is the boundary of the $AdS$ space.
Said another way, we can write the code rate (with $\textbf{$S_b$}=0$) to be
\begin{equation}
    \frac{k}{n}=e^{(r_{in}-r_{out})/L},
\end{equation} 
where $k$ and $n$ are the number of logical and physical qubits respectively expressed in terms of lengths of inner and outer boundaries.
With the introduction of mixed state in the bulk, $r_{in}\rightarrow r'_{in}$, implying that the code rate depreciates with $k\rightarrow k'$ ($k'>k)$. As long as $k'<<n$, the is a reasonable error correcting scheme.

\section{The Holographic Sierpinski Code In $AdS_4/CFT_3$}
\begin{figure}[h!]
		\centering
		\includegraphics[width=0.5\textwidth]{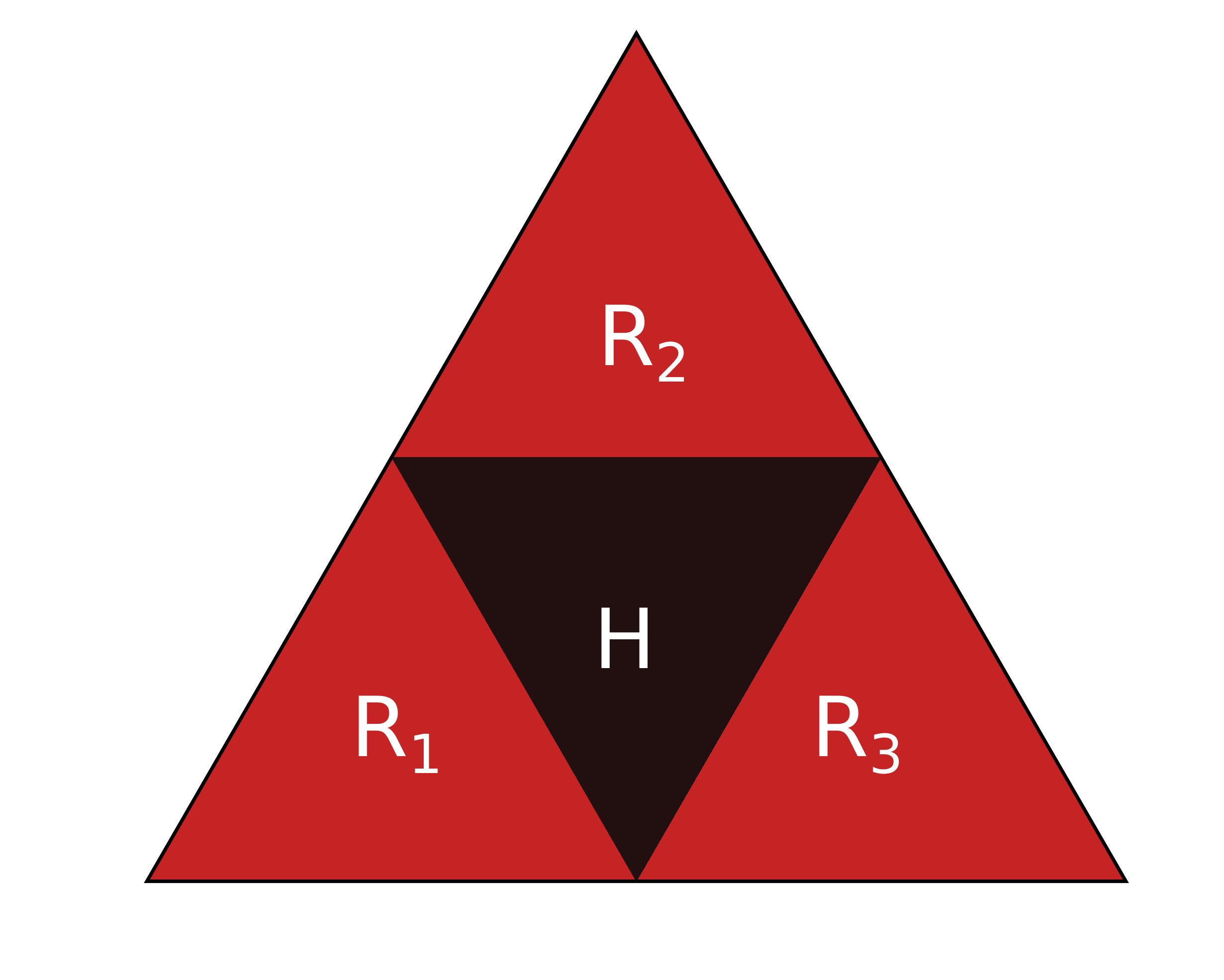}
	\caption{We divided the full triangle $R$ into four triangular regions and punched a hole $H$ out of it. The remaining portion of the boundary are the union of three triangles $R_1$, $R_2$ and $R_3$.}
	\label{fig:triangle-level1}
\end{figure}

An example of uberholography in $AdS_4/CFT_3$ with a Sierpinski triangle shaped boundary is constructed in \cite{bao2022}. 
We will briefly review this. Consider the boundary Sierpinski triangle at level 0 (without holes) which is an equilateral triangle having sides of length $l_0$. At level 1, we divide this triangle $R$ into four smaller triangles and make a hole $H$ by taking away the equilateral triangle in the center of side $l_1=\frac{l_0}{2}-\epsilon$ where $\epsilon>0$ is extremely small compared to $l_0$. It suffices to take the length of the remaining triangles $R_1$, $R_2$ and $R_3$ to be $\frac{l_0}{2}$(see Fig. \ref{fig:triangle-level1}).

\begin{figure}[h!]
		\centering
		\includegraphics[width=0.5\textwidth]{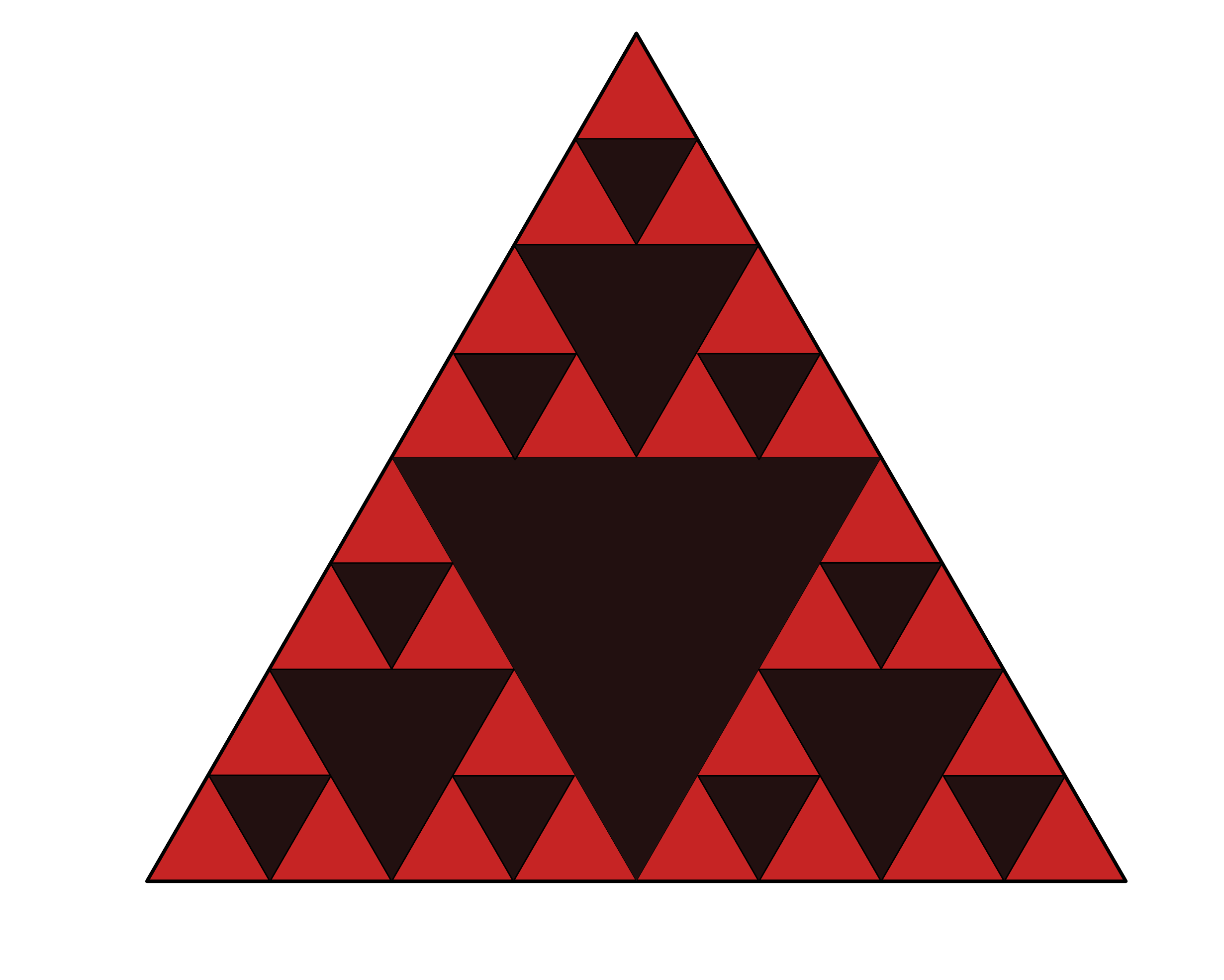}
	\caption{The Sierpinski Triangle. The fractal is constructed by removing triangular holes (shaded black) of decreasing size from the big triangle.}
	\label{fig:sierpinski-triangle}
\end{figure}

The punching of holes is re-iterated (see Fig. \ref{fig:sierpinski-triangle}). At level $m$, the $3^m$ small triangles are taken to have a side of length $l_m=l_0/2^m$, while the length of triangles forming holes ranges from  $l_0/2-\epsilon$ to $l_0/2^m-\epsilon$. This construction is more subtle in the sense that the connected phase and disconnected phase have equal RT surface area at leading order at every level of hole iteration in the $\epsilon=0$ limit. It is shown in \cite{tonni2015} that the minimal surface associated with a regular polygon with sharp corner picks up a corner contribution. For the case of a boundary subregion $A$ in the shape of equilateral triangle of side $L$, the area of RT surface is given by
\begin{equation}
    |\chi_A|=\frac{3L}{a}-6b\left(\frac{\pi}{3}\right)\log{\left(\frac{3L}{a}\right)}
\end{equation}
where $a$ is the short-distance cutoff and $\chi_{A}$ is the RT surface associated with the boundary subregion $A$. Here $b(\frac{\pi}{3})$ is a regulator-independent coefficient that depends only on the opening angle (in this case $\pi/3)$ defined in \cite{tonni2015}. See \cite{myers2015} for illuminating details about this coefficient.
Recalling that in the absence of a bulk entropy, one arrived at the following inequality at level 1
\begin{equation}
    |\chi_{R}|+|\chi_{H}|\leq |\chi_{R_1}|+|\chi_{R_2}|+|\chi_{R_3}|,
\end{equation}
where the boundary is divided as shown in Fig. \ref{fig:triangle-level1}. 

Solving the critical case where this inequality is saturated under reasonable approximations discussed in \cite{bao2022}, we get a lower bound on the value of $\epsilon$
\begin{equation}
    \epsilon\geq 2ab\left(\frac{\pi}{3}\right)\log{\frac{3l_0}{4a}}.
\end{equation}
For a generic level $m$, $\epsilon$ depends on $m$ and now in general it would also depend on the bulk entropy $\textbf{$S_b$}$.
This should be contrasted with the construction in $AdS_3/CFT_2$, where in the absence of bulb entropy $\textbf{$S_b$}$, the critical value of $r$ did not depend on the level of $m$. The $m$-dependence of $r$ was introduced by the presence of bulk entropy $\textbf{$S_b$}$ on which $r$ depended as well (as illustrated in Fig. \ref{fig:vary-m}).

\subsection{Reconstruction Wedge of the Sierpinski Triangle}
\begin{figure}[h!]
		\centering
		\includegraphics[width=0.5\textwidth]{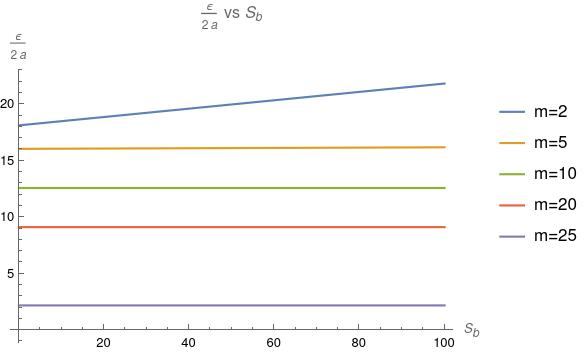}
	\caption{$\epsilon$ vs $S_b$ plot for various $m$. We can see the linear dependence on $S_b$ for small $m$. At large $m$, dependence on $S_b$ is suppressed. The limiting case is thus $\epsilon=a\left(2b(\pi/3)\log{3}\right)$. We have set $l_0/a=1,000,000,000$, $b(\pi/3)=1$ and $2G_N=1$ for simplicity.}
	\label{fig:4d-vary-m}
\end{figure}

In the presence of bulk entropy \textbf{$S_b$}, at level 1, we want to solve the inequality
\begin{equation}\label{qecc-condition}
    |\chi_{R}|+|\chi_{H}|+4G_N\textbf{$S_b$}\leq |\chi_{R_1}|+|\chi_{R_2}|+|\chi_{R_3}|,
\end{equation}
where $G_N$ is the Newton's constant in 3+1 dimensions.
This gives us a lower bound on $\epsilon$
\begin{equation}
    \label{eps-level-1}
    \epsilon\geq a\left[\frac{4G_N\textbf{$S_b$}}{3}+2b\left(\frac{\pi}{3}\right)\log{\left(\frac{3l_0}{4a}\right)}\right].
\end{equation}
As we see $\epsilon$ scales linearly in \textbf{$S_b$} and we can work in the small $\epsilon$ limit, as long as the product of length cutoff $a$ and \textbf{$S_b$} remains small. Moving on, we are interested in the case, where there are large number of iterations of hole punching.
After $m$ iterations, the connected phase has an area

\begin{equation}\label{conn-m-iters}
\begin{split}
|\chi_{R'}|_{c.} = & |\chi_{R}|+|\chi_{H_1}|+ 3|\chi_{H_2}|+ 3^2 |\chi_{H_3}|+ \dots + 3^{m-1} |\chi_{H_m}|  \\
 = & |\chi_{R}|+ \sum_{j=1}^{m}3^{j-1}|\chi_{H_j}| \\
 = & \left[\frac{3l_0}{a}+ \sum_{j=1}^m 3^{j-1}\frac{3l_0}{2^j a}\right] -\frac{3\epsilon}{a}\sum_{j=1}^m 3^{j-1}\\ &
-6 b\left(\frac{\pi}{3}\right) \left[\log{\left(\frac{3l_0}{a}\right)}+ \sum_{j=1}^m 3^{j-1}\log\left({\frac{3l_0}{2^j a}}\right)\right]\\
& -6 b\left(\frac{\pi}{3}\right)\left[\sum_{j=1}^m 3^{j-1}\log\left(1-\frac{2^j\epsilon}{l_0}\right) \right],
\end{split}
\end{equation}
while the area of the disconnected phase is
\begin{equation}\label{discon-m-iters}
    |\chi_{R'}|_{disc.}=3^m \left(\frac{3l_0}{2^m a}\right)-3^m 6b\left(\frac{\pi}{3}\right)\log{\left(\frac{3l_0}{2^m a}\right)}.
\end{equation}
Introducing an entropic state in the bulk with von Neumann entropy \textbf{$S_b$}, the bound on $\epsilon$ comes out to be
\begin{equation}
    \begin{aligned}
    \epsilon \geq 2a\left[\frac{2G_N \textbf{$S_b$}}{3^{m+1}}+b\left(\frac{\pi}{3}\right)\left(\log{\left(\frac{3l_0}{2^m a}\right)}-\frac{m(m+1)}{3^m}\log{2}\right)\right].
    \end{aligned}
\end{equation}
As long as we are in the regime where at level $m$, the quantity ($\frac{l_0}{2^m}$) is of the order of lattice spacing '$a$' (the ratio $\frac{l_m}{a}=\mathcal{O}(1)$), in the very large $m$ limit, the linear dependence on the bulk entropy \textbf{$S_b$} is suppressed by a very large factor $3^{m+1}$. Effectively, in this limit, one can drop the first and the third term, arriving at
\begin{equation}
    \epsilon=2ab\left(\frac{\pi}{3}\right)\log{\left(\frac{3l_0}{2^m a}\right)},
\end{equation}
which is state-independent.
However, it is worth mentioning that as we make the length of the side of the triangle smaller than this cutoff length '$a$', there is a breakdown of the entanglement wedge. Such a case would mean that the cutoff could be bounded below by a negative number and is thus unphysical. This is perhaps a relic of the untrustworthiness of theories below cutoffs.
As we are talking about the Sierpinski triangle as boundary to $AdS_4$, it is reasonable to assume $l_0$ is very large and short-distance cutoff '$a$' is very small, one can always choose sufficiently large $m$ to allow for state-independent bulk reconstruction (see Fig. \ref{fig:4d-vary-m}).

\section{\label{sec:conclusion}Conclusion}
We see that the robustness of the holographic quantum error correcting code against fractal noise persists even in finite temperature states of the CFT, or equivalently in black hole backgrounds. This is a priori surprising, as one might have thought that the black hole horizon size would have served as a finite cut-off to the uberholography generalization to higher dimensions. We believe that any purely IR contribution (e.g, a black hole in the center) becomes entropically negligible when the fractal contributions of (infinitely) many holes are considered, rendering the property of state-independence. A recent work \cite{uberholo2022} that appeared at the time of completion of this manuscript discusses uberholography in Cantor-set like erasures in higher dimensions. It also discusses the finite temperature effects on uberholography. This work is complementary to our work and tells a similar story in a different language.\footnote{To study finite temperature CFT, the work in \cite{uberholo2022} has deformed the bulk metric to a BTZ blackhole while we continue to work in empty AdS-metric. When the entanglement wedge boundaries are very far from the blackhole horizon in the bulk, in that limit, the metric near the boundary is near-vacuum. Figure (1) \cite{uberholo2022} show similar trend to our figure (3).}

Given the continued robustness of uberholographic codes to fractal noise, the logical next step would be to determine the precise features of the holographic quantum error correcting code that permit its robustness against this form of error, and to port a novel code with such features into the realm of quantum computing. We will reserve both of these studies for future work.

\begin{acknowledgments}
All authors would like to thank Chris Akers for useful discussions and acknowledge the referee for suggestions. N.B. is supported by the Department of Energy under grant number DESC0019380, and is supported by the Computational Science Initiative at Brookhaven National Laboratory, and by the U.S. Department of Energy QuantISED Quantum Telescope award. J.N. is supported by a Graduate Assistantship from the Department of
Physics, Northeastern University.
\end{acknowledgments}



\nocite{*}

\bibliography{apssamp}

\end{document}